\documentclass{sf2a-conf}
\usepackage{graphicx}
\usepackage{natbib}
\begin{document}

\TitreGlobal{SF2A 2008}

\title{The timescale for giant planet formation~: constraints from the
rotational evolution of exoplanet host stars.}
\author{Bouvier, J.}\address{Laboratoire d'Astrophysique,
  Observatoire de Grenoble, Universit\'e J. Fourier, CNRS, BP 53,
  38041 Grenoble, Cedex 9, France}
\runningtitle{The rotational evolution of exoplanet host stars }
\setcounter{page}{237}

\index{Bouvier, J.}

\maketitle
\begin{abstract}
The timescale over which planets may form in the circumstellar disks of
 young stars is one of the main issues of current planetary formation
 models.  We present here new constraints on planet formation
 timescales derived from the rotational evolution of exoplanet host
 stars.
\end{abstract}
%
\section{Introduction}

The time it takes to form giant gaseous planets in the circumstellar
disks of young stars is still a poorly constrained parameter. On the
theoretical side, models predict planet formation timescales in the
range from $\sim$1~Myr to 10~Myr, depending on the processes at work
(e.g. Ida \& Lin 2004; Alibert et al. 2005; Guillot \& Hueso 2006;
Lissauer \& Stevenson 2007). On the observational side, protoplanetary
disk lifetimes, as measured by the decay of either infrared excess
(dust) or line emission (gas) in pre-main sequence stars, appear to
vary from star to star, in the range from $\leq$1~Myr up to about
10~Myr (e.g. Lawson et al. 2004; Hillenbrand et al. 2005; Jayawardhana
et al. 2006; Meyer et al. 2007). Why do some stars dissipate their
disk on very short timescales while other retain their disk up to
$\sim$10~Myr ? Is rapid disk dissipation the result of prompt planet
formation in the disk ? Or, on the contrary, are long-lived disks
required to allow for planet formation ?

Indirect clues may be gained by investigating the imprint the planet
formation process may have left on the properties of exoplanet host
stars. Israelian et al. (2004) reported that solar-type stars with
massive planets are more lithium depleted than their siblings without
detected massive planets, a result recently confirmed by Gonzalez
(2008). We investigate here whether enhanced lithium depletion in
exoplanet host stars may result from their specific rotational
history, which in turn is tightly coupled to the evolution of their
circumstellar disk during the pre-main sequence. In this way, we
attempt to relate giant planet formation to lithium abundances,
angular momentum evolution, and disk lifetimes.

\section{The rotational evolution of solar-mass stars} 


Figure~1 shows models we developped to investigate the rotational
evolution of solar-type stars, from their birth up to the age of the
Sun. The models dicussed here were originally developped by Bouvier et
al. (1997) and Allain (1998). The rotational evolution of solar-mass
stars is driven by a number of physical processes acting over the
star's lifetime. During the early pre-main sequence (PMS), the star is
magnetically coupled to its accretion disk (cf. Bouvier et
al. 2007). As long as this interaction lasts, the star is prevented
from spinning up (in spite of contraction) and evolves at constant
angular velocity (Matt \& Pudritz 2005). The disk lifetime, a free
parameter of the model, thus dictates the early rotational evolution
of the star. When the disk eventually dissipates, the star begins to
spin up as it contracts towards the zero-age main sequence
(ZAMS). Depending on the initial velocity and disk lifetime, a wide
range of rotation rates can be obtained on the ZAMS (Bouvier et
al. 1997). The lowest initial velocities and longest disk lifetimes
result in the slowest rotation rates on the ZAMS. On the opposite,
high initial velocities and/or short disk lifetimes lead to fast
rotation on the ZAMS. Finally, as the stellar structure stabilizes on
the ZAMS, at an age of about 40 Myr for a solar-mass star, the braking
by a magnetized wind becomes the dominant process and effectively
spins the star down on the early main sequence (MS). As the braking
rate scales with surface velocity (Kawaler 1988), fast rotators are
spun down more efficiently than slow ones, and this leads to a rapid
convergence towards uniformly slow rotation by the age of the
Sun. Indeed, after a few Gyr, the surface rotational velocity of
solar-type stars has lost memory of the past rotational history.

Internal differential rotation is an important additional parameter of
the model. We consider here a radiative core and a convective envelope
that are each in rigid rotation, but whose rotation rate may differ
(Allain 1998). We therefore introduce a coupling timescale between the
inner radiative zone and the outer convective envelope, $\tau_c$,
which measures the rate of angular momentum transfer between the core
and the envelope (MacGregor \& Brenner 1991). A short coupling
timescale corresponds to an efficient core-envelope angular momentum
transport and, as a consequence, little internal differential
rotation. On the opposite, a long coupling timescale leads to the
developement of a large rotational velocity gradient between the core
and the envelope. This model parameter, $\tau_c$, governs internal
differential rotation, and is therefore expected to be of prime
importance for rotationally-induced mixing and associated lithium
depletion during the evolution of solar-type stars.

   \begin{figure}
   \centering
  \includegraphics[width=0.8\textwidth]{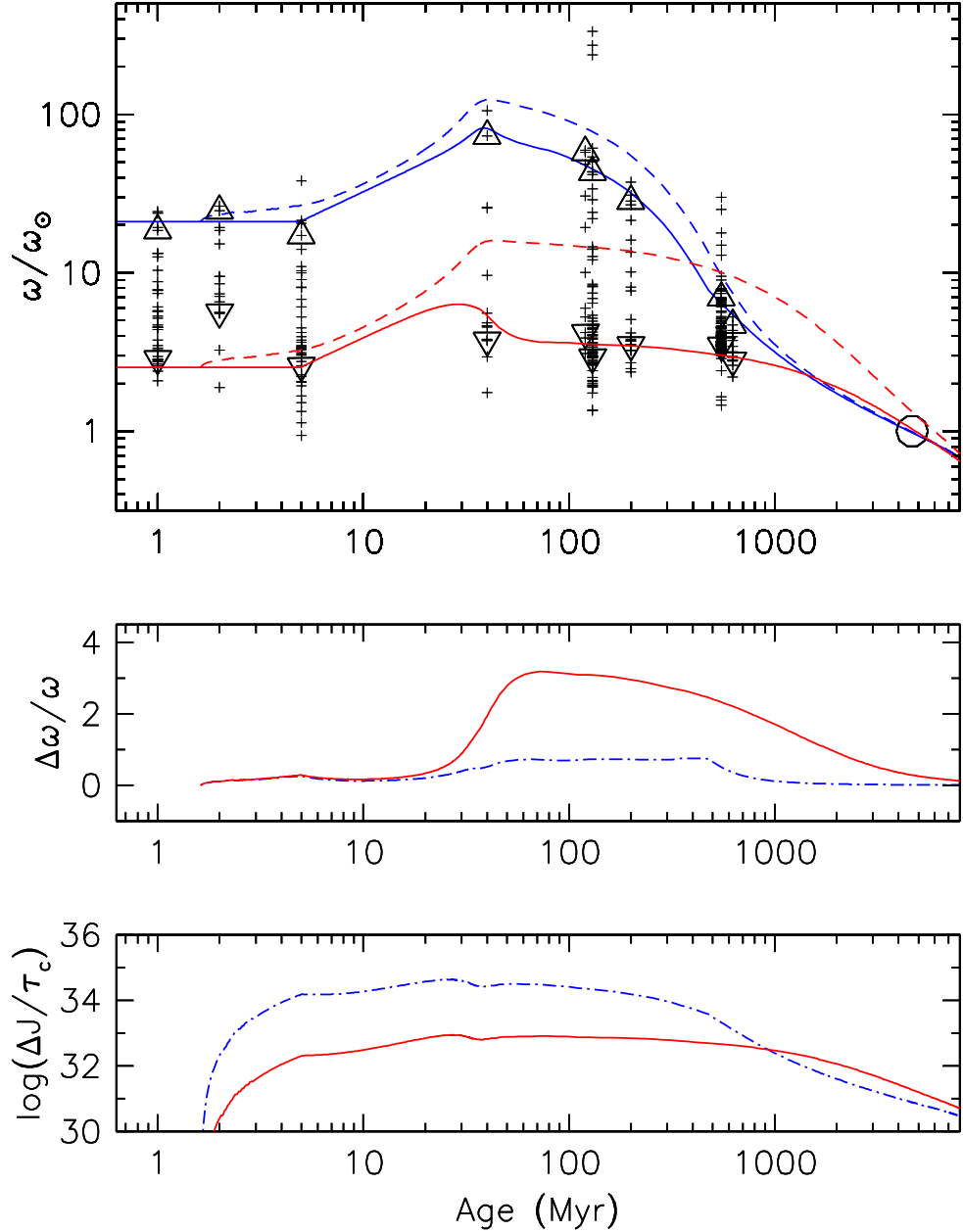}
   \caption{Rotational models for slow and fast solar-mass
rotators. {\it Data~:} The 10th and 75th percentiles of the observed
rotational period distributions of solar-type stars
(0.8-1.1~M$_\odot$) were converted to angular velocity and are plotted
as direct and inverted triangles as a function of time. Individual
measurements of rotational periods converted to angular velocities are
also shown in order to illustrate the statistical significance of the
various samples. {\it Models~:} Rotational evolution models are shown
for slow and fast 1~M$_\odot$ rotators. For each model in the upper
panel, surface rotation is shown as a solid line, and the rotation of
the radiative core by a dashed line. With a core-envelope coupling
timescale $\tau_c$ of only 10~Myr, little differential rotation
develops in fast rotators. In contrast, the 100~Myr core-envelope
coupling timescale in slow rotators results in a large velocity
gradient at the base of the convective zone. A disk lifetime of 5~Myr
is assumed for both models. {\it Lower panels :} The velocity shear at
the base of the convective zone
($\omega_{rad}-\omega_{conv})/\omega_{conv}$, and the angular momentum
transport rate $\Delta J/\tau_c$ ($g cm^2 s^{-2}$) from the core to
the envelope are shown for slow (solid line) and fast (dotted-dashed
line) rotators.}
              \label{model}%
    \end{figure}

The models are confronted to the observed rotation rates of solar-type
stars at various ages (e.g. Irwin et al. 2008). We aim here at
reproducing the lower and upper envelopes of the observed rotational
distributions, in order to contrast the evolution of slow and fast
rotators and relate it to lithium depletion. A model for fast rotators
is compared to observations in Fig.~\ref{model}. Starting from an
initial period of 1.2~d, the star remains coupled to its disk for
5~Myr, then spins up to a velocity of order of 160~km~s$^{-1}$ on the
ZAMS, and is eventually spun down by a magnetized wind on the MS to
the Sun's velocity. The model reproduces reasonably well the PMS spin
up and the rapid MS spin down observed for fast rotators between 5 and
500~Myr. In order to reach such an agreement, the core-envelope
coupling timescale has to be short, $\tau_c \sim $ 10~Myr, which
implies little internal differential rotation in fast rotators.

Fig.~\ref{model} also shows a model for slow rotators. The initial
period is 10~d and the star-disk interaction lasts for 5~Myr in the
early PMS. As the star approaches the ZAMS, both the outer convective
envelope and the inner radiative core spin up. Once on the ZAMS, the
outer envelope is quickly braked, while the core remains in rapid
rotation. This behaviour results from an assumed weak coupling between
the core and the envelope, with $\tau_c\sim$ 100~Myr. On the early MS,
the rapidly-rotating core transfers angular momentum back to the
envelope, which explains the nearly constant surface velocity over
several 100~Myr in spite of magnetic braking. We thus find that a long
core-envelope coupling timescale is required to account for the
observed rotational evolution of slow rotators, which implies the
developement of a large velocity gradient at the core-envelope
boundary.

\section{Lithium depletion, rotation, and the lifetime of
  protoplanetary disks}

The modeling of the rotational evolution of solar-type stars seems to
imply that internal differential rotation is much larger in slow
rotators than in fast ones. This should have a strong impact on
lithium abundances, as the efficiency of rotationally-induced lithium
burning is expected to scale with differential rotation (Zahn
2007). This model prediction is supported indeed by measurements of
lithium abundances in the Pleiades open cluster, at an age of
100~Myr. Soderblom et al. (1993) found that rapidly rotating
solar-type stars in the Pleiades exhibit higher lithium abundances
than slow rotators, which indicates that lithium depletion already
takes place during the PMS/ZAMS, and is more pronounced in slow than
in fast rotators.

Different rotational histories may thus be reflected in the lithium
abundance pattern of mature solar-type stars, leading to a {\it
dispersion\/} of lithium abundances at a given age and mass, long
after the circumstellar disks have disappeared. The models above
suggest that enhanced lithium depletion is associated to low surface
rotation on the ZAMS. Then, the fact that mature solar-type stars with
massive exoplanets are lithium-depleted compared to similar stars with
no planet detection seems to indicate that massive exoplanet hosts had
slow rotation rates on the ZAMS.

Why were massive exoplanet host stars slow rotators on the ZAMS ?  Two
main parameters dictate the rotation rate at the ZAMS : the initial
velocity and, most importantly, the disk lifetime. For a given disk
lifetime, the lower the initial velocity, the lower the velocity on
the ZAMS. Conversely, for a given initial velocity, the longer the
disk lifetime, the lower the velocity on the ZAMS. This is because the
magnetic star-disk interaction during the PMS is far more efficient
than solar-type winds in extracting angular momentum from the star
(Bouvier 2007; Matt\& Pudritz 2007). Disk lifetimes varying from star
to star in the range 1-10 Myr are required to account for the
distribution of rotational velocities on the ZAMS (Bouvier et
al. 1997).  Statistically, however, the slowest rotators on the ZAMS
are expected to be the stars who had initially low rotation rates {\it
and the longest-lived disks}. An initially slowly-rotating star with a
short-lived disk would strongly spin up during the PMS and reach the
ZAMS as an intermediate or fast rotator.

Long-lived disks thus appear as a necessary condition for massive
planet formation and/or migration on a timescale $\geq$5~Myr. Long
lasting disks may indeed be the common origin for slow rotation on the
ZAMS, lithium depletion and massive planet formation.  Interestingly
enough, the Sun hosts massive planets. Even though the solar system
gaseous planets are located further away from the Sun than massive
exoplanets are from their host stars, the Sun is strongly lithium
deficient. According to the scenario outlined above, the Sun would
thus have been a slow rotator on the ZAMS.

\section{Conclusions}

Based on what we currently know of the rotational properties of young
stars, of the lithium depletion process in stellar interiors and of
the angular momentum evolution of solar-type stars, it seems likely
that the lithium-depleted content of massive exoplanet host stars is a
sequel to their specific rotational history. This history is
predominantly dictated by star-disk interaction during the pre-main
sequence. Rotationally-driven lithium depletion in exoplanet host
stars can be at least qualitatively accounted for by assuming
protoplanetary disk lifetimes of order of 5-10 Myr. Such long-lived
disks may be a necessary condition for planet formation and/or
migration around young solar-type stars, at least for the class of
giant exoplanets detected so far. A full account of this work is given
in Bouvier (2008).



\end{document}